\documentclass[preprint,12pt]{elsarticle}
\usepackage[pdftex,
			colorlinks={true},
			pdfstartview={FitH}
            ]{hyperref}

\usepackage{graphicx}

\usepackage{paralist}
\usepackage{longtable}
\usepackage{array}
\usepackage{epstopdf}
\usepackage{qtree}

\usepackage{amssymb}

\begin{document}
\qtreecenterfalse
\begin{frontmatter}



\title{Modeling Website Workload Using Neural Networks}


\author{Yasir Shoaib}
\ead{yasir.shoaib@ryerson.ca}
\author{Olivia Das}
\ead{odas@ee.ryerson.ca}
\address{Department of Electrical and Computer Engineering, Ryerson University, Toronto, ON M5B 2K3}

\begin{abstract}
In this article, artificial neural networks (ANN) are used for modeling the number of requests received by 1998 FIFA World Cup website. Modeling is done by means of time-series forecasting. The log traces of the website, available through the Internet Traffic Archive (ITA), are processed to obtain two time-series data sets that are used for finding the following measurements: requests/day and requests/second. These are modeled by training and simulating ANN. The method followed to collect and process the data, and perform the experiments have been detailed in this article. In total, 13 cases have been tried and their results have been presented, discussed, compared and summarized. Lastly, future works have also been mentioned.
\end{abstract}


\end{frontmatter}














\section{Introduction}
Forecasting the arrival rate of requests to a website helps the web developer and provider prepare ahead and accordingly meet desired performance objectives. If the workload intensity --- which could be expressed as incoming request rate or as number of concurrent users in the system \cite[p.~19]{2011_YasirShoaibMAScThesis}\cite[p.~58]{1984_QSP} --- can be predicted, then necessary computing resources could be made available, thereby preventing device saturation that cause long end-to-end response times.

For the purposes of predicting the future workloads, in this article, artificial neural networks (ANN) are used to model the workload of 1998 FIFA World Cup website, by means of time-series forecasting. Actual log traces of the World Cup site (1.35 billion requests) \cite{WorldCup98Traces} are available from The Internet Traffic Archive (ITA) \cite{LogTraces}. Other noteworthy traces available through the ITA website include logs of NASA website (about 3.46 million requests) \cite{NASAHTTP} and EPA webserver (47,748 requests) \cite{EPAHTTP}. Furthermore, ITA provides tools to help read and process the logs. Due to availability of the source code, custom enhancements to these tools are possible.

The main reasons for choosing of neural networks for the modeling are their ability to approximate non-linear behavior and them being ``data-driven'' \cite{1998_ForecastingANN,2001_NNShortTermLoadForecast}. It is best to assume that for real examples (just like the website data as in our case), the inputs and the outputs have a non-linear relationship, instead of assuming a linear relationship as considered by traditional approaches \cite{1998_ForecastingANN}. Furthermore, with the data available it would be understandable to directly feed it to the ANN and obtain the results of the function approximation rather than judge beforehand about the nature of the functional relationships. This is because ANN can implement ``nonlinear modeling'' \cite[p.~36]{1998_ForecastingANN} without beforehand knowledge of the input and output relationships.

Previously, just like this article, similar works related to web traffic modeling and prediction using ANN have been done by Prevost et al. \cite{2011_PredictionCloudNN} and Chabaa et al. \cite{2010_IdentPredictInternetTraffic}. Prevost et al. \cite{2011_PredictionCloudNN} obtain the trace logs of NASA website \cite{NASAHTTP} and EPA web server \cite{EPAHTTP}, and use ANNs and regressive linear prediction to guess the next few seconds of request rate, where step-ahead intervals range in-between 1 second and 90 seconds. For measuring performance mean-squared error (MSE) and root mean-squared error (RMSE) errors were calculated and the results of the RMSE were shown. Chabaa et al. \cite{2010_IdentPredictInternetTraffic} model 1000 data points using different training algorithms (including Levenberg-Marquardt algorithm (LM) \cite{2013_MatlabLM}) on multi-layer perceptron (MLP) neural network and compare their performances. In contrast, this article focuses only on LM algorithm for training purposes. Few notable differences between the previous papers and this article is that 1998 FIFA World cup website logs depict a busier site and has lot more requests, in particular 1.35 billion requests averaging to about 15.3 million requests/day during a period of 88 days, when compared to a total about 3.46 million requests of the NASA website and 47,748 requests of EPA server. Furthermore, in this article, both requests/seconds and requests/day data sets are modeled using ANN. Here the focus is toward one-step ahead prediction, although for one case two-step ahead prediction is also performed.

Saripalli et al. \cite{2011_LoadPredictionHotSpotDetection} have also presented their work relating to workload prediction; however, their approach relies on a two step process, first of which is associated with tracking of the workload, followed by the prediction step. Although, our approach does not use workload tracking, ANN could be used for prediction purposes in second step of the aforementioned paper. 

Giang et al. \cite{2013_NeuralWebServerWorkload} in their paper compare five neural network based models based on their performance for the forecasting of hourly HTTP workload of a commercial website. The performance is measured through Mean Absolute Percentage Error (MAPE). Based on the study, it is seen that non-linear autoregressive with exogenous input (NARX) neural networks are able to predict the workload best. One difference between their approach and ours is that they also use previous 6 values of the workload as the input to perform the prediction, where each input value is separated by 24 hours, i.e. past values at time t, t-24, t-48, etc. are used to predict the value at time t+24.

In this article, MATLAB Neural Network Toolbox \cite{2012_MatlabNNGS,2012_MatlabNNUG} has been used for creating of neural network models and simulating them. Two data sets --- \textit{day-requests} and \textit{epoch-requests}\footnote{Epochs are integers and in this article refer to seconds passed since the epoch time, Jan 1, 1970 \cite{2011_timeManPage}. Therefore, \textit{epoch-requests} data set is used for finding requests/second}  --- are input for training the neural networks. The first data set is used for predicting of requests/day and the second for prediction of requests/second. In total 13 cases have been tried and their mean-squared errors (MSE) presented in this article. The first 11 cases correspond for \textit{day-requests} data and the remaining two correspond to \textit{epoch-requests}. Through this article the applicability of ANN in modeling and forecasting of website workload intensity has been demonstrated; this is the main contribution of this work.

The structure of the article is as follows. In section \ref{sec:FIFAData} the FIFA World Cup website data and data collection process are discussed. Section \ref{sec:ANN_MATLAB} provides a brief introduction to ANN and MATLAB. Section \ref{sec:Method} describes the methods followed. Section \ref{sec:Results} presents the results. Section \ref{sec:Conclusions} concludes the article. 

\section{1998 World Cup Website Data}
\label{sec:FIFAData}
The 1998 World Cup commenced on June 10, 1998 and finished on July 12, 1998 and was played by 32 teams, totalling 64 matches \cite{1998_FIFAArchive,2000_1998WorldCupWorkloadCharacter}. It was hosted by France \cite{2000_1998WorldCupWorkloadCharacter}, also the team to snatch the World Cup that year, winning against the the defending champions Brazil in a final score of 3-0 \cite{1998_FranceDefeatBrazil}. 

The \url{http://www.france98.com}\footnote{The URL \url{http://www.france98.com} does not appear to correspond to the 1998 FIFA World Cup anymore. Interested readers, may visit the 1998 FIFA Archive available at \url{http://www.fifa.com/worldcup/archive/edition=1013/index.html}, which includes photographs and archived information of the 1998 World Cup.} website for the 1998 World Cup, was hosted from May 6, 1997 to serve its users with live match scores, team and player stats, match photographs, interviews and much more \cite{2000_1998WorldCupWorkloadCharacter}. The site's trace logs of received requests were collected from April 30, 1998 to July 26, 1998, a period of 88 days comprising of total 1,352,804,107 requests \cite{WorldCup98Traces}. Traces were obtained from 33 web servers that resided in four locations: one location in Paris, France and the rest three spread across USA \cite{WorldCup98Traces}. The ITA website \cite{LogTraces}, which hosts the logs has 92 days of trace log data, the first four days of which are empty --- representing April 26, 1998 as day 1 --- and used as a filler to help with identifying weekdays. Each day's log data is further divided into files of maximum 7 million requests, thereby limiting the file size to within 50 MB and causing one day log data to be associated with multiples files \cite{WorldCup98Traces}. For example, day 38 is divided into two files: \texttt{wc\_day38\_1.gz} (6,999,999 requests) and \texttt{wc\_day38\_2.gz} (188,042 requests). There are in total 249 binary files, which need further processing to read their contents. For this purpose, the 1998 FIFA log site \cite{WorldCup98Traces} includes the following three tools useful here: \texttt{read}, \texttt{recreate} and \texttt{checklog}. The \texttt{read} tool aids in counting of number of requests in each file, the \texttt{recreate} tool displays the log contents after converting them from binary, and \texttt{checklog} presents the request statistics from the information in the binary files. For readers who are interested in an in-depth discussion and analysis of the 1998 World cup website workload may refer to article by Arlitt and Jin \cite{2000_1998WorldCupWorkloadCharacter}.

\begin{figure}[h!t!b]
\centering
\includegraphics[keepaspectratio,viewport=0 0 1405 602,clip,width=\textwidth]{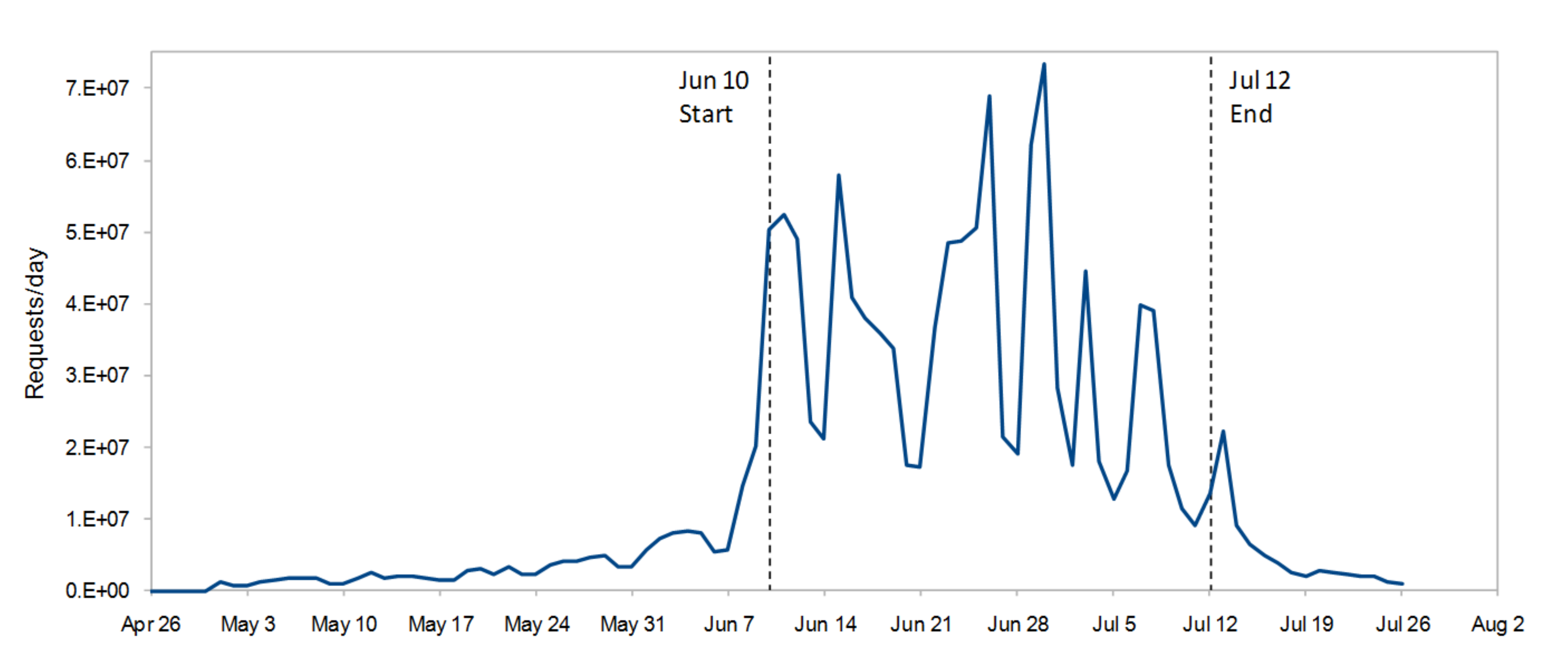}
\caption{Requests/day graph showing the fluctuating workload received by the 1998 World cup website before, during and after the World cup matches were held. The trace logs were collected from April 30, 1998 although the graph includes empty values from April 26, 1998 to help identify weekdays. The world cup started on June 10, 1998 and ended on July 12,1998. The highest workload was witnessed on day 66 (June 30, 1998) accounting for a total of 73,291,868 requests.}
\label{fig:RequestsDayGraph}
\end{figure}

\figurename~\ref{fig:RequestsDayGraph} shows the fluctuating requests/day graph, depicting the requests received by the website. This graph is derived by plotting the day (x-axis) and requests (y-axis) columns of the \textit{day-requests} data set, which was extracted from the trace logs. As seen, the popularity of the website increased as approaching the beginning of the World cup and decreased at a quick rate after the end of matches. The highest workload was witnessed on day 66 (June 30, 1998) accounting for a total of 73,291,868 requests. 

\begin{figure}[h]
\centering
\includegraphics[keepaspectratio,viewport=0 0 1231 409,clip,width=\textwidth]{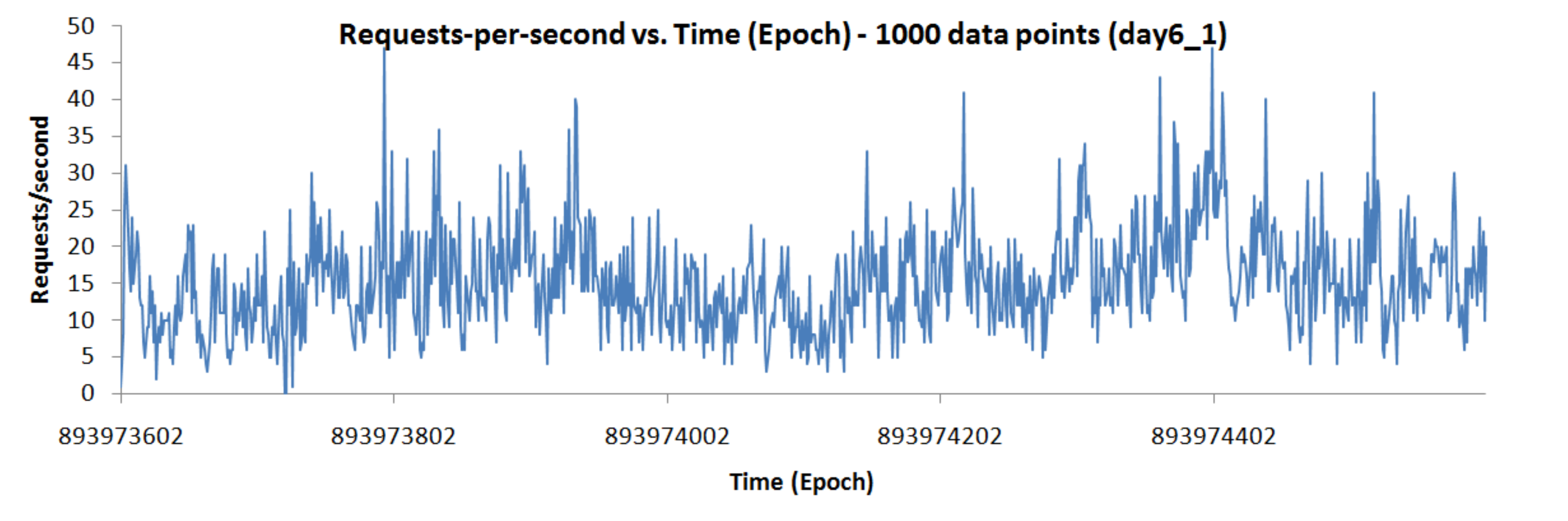}
\caption{Requests/second vs. epoch graph for day 6 (first 1000 data points)}
\label{fig:RequestEpoch6}
\end{figure}

\begin{figure}[h]
\centering
\includegraphics[keepaspectratio,viewport=0 0 1231 409,clip,width=\textwidth]{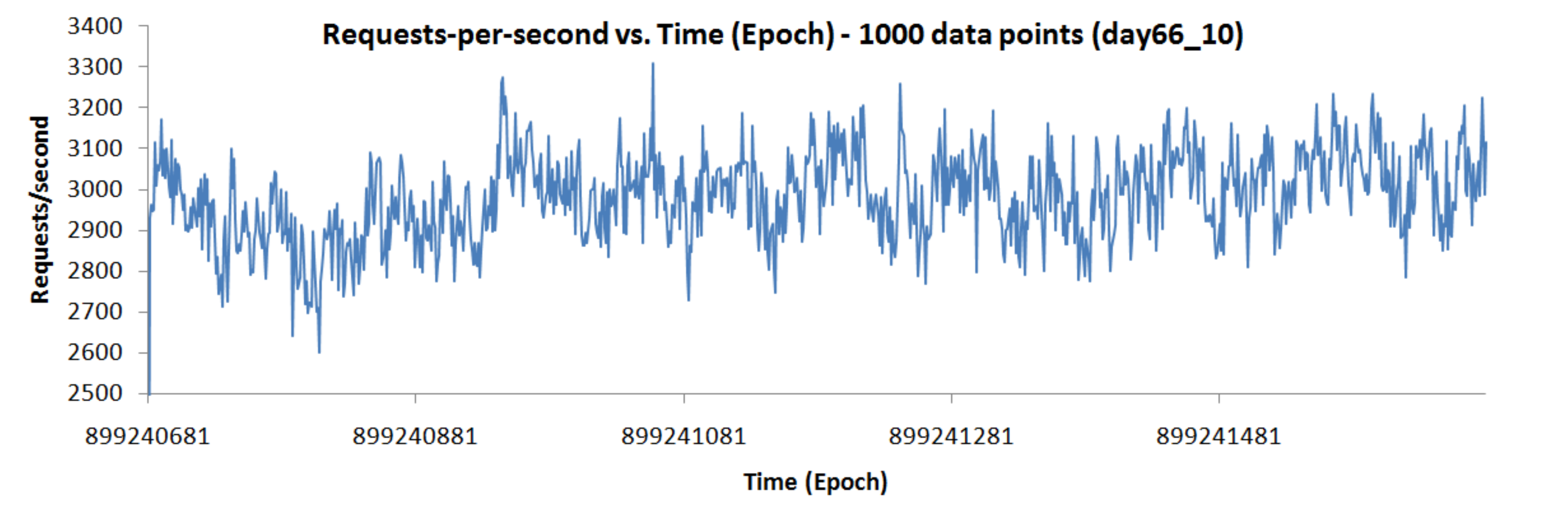}
\caption{Requests/second vs. epoch graph for day 66-part10 (first 1000 data points)}
\label{fig:RequestEpoch66_10}
\end{figure}

\figurename~\ref{fig:RequestEpoch6} and \figurename~\ref{fig:RequestEpoch66_10} show the requests/second vs. epoch graph for the first 1000 points of day 6 and day 66-part10\footnote{As discussed earlier in this section, each days data is divided into files of maximum 7 million requests because of which day 66 log files are divided into 11 parts. For this article we use the 10th file for our purposes.} files, respectively. It is key to realize the distinction between day 6 and the day 66 files in general. Day 6 have fewer requests/second with maximum rate around 50 requests/second however day 66 has a higher request rate with maximum rate around 3300 requests/second with a higher fluctuation per second as seen from the graph. These graphs have been derived by processing the log traces to obtain \textit{epoch-requests} data set and then using the latter for graphing. On this note, the following: section \ref{subsec:dataCollection} and section \ref{subsec:dataFormat} describe the data collection and data format of the data sets used in this article.

\subsection{Data Collection}
\label{subsec:dataCollection}
To collect data and generate the two required data sets: \textit{day-requests} and \textit{epoch-requests}, the first task was to download the log trace files from the ITA website. \texttt{wget} utility, which is available on Linux-based operating systems, was used for downloading the files; however, any browser may be used to easily click and download the files. The downloaded file size amounted to about 9 GB and the files were placed in a Log Trace folder on the hard drive. Afterwards, the \texttt{read} program was run against each of the 249 binary files in the Log Trace folder to count the number of requests in each and subsequently use the information to find number of requests/day for each day, thereby generating the \textit{day-requests} data set. 

\begin{figure}[h!t!b]
\centering
\includegraphics[keepaspectratio,viewport=0 0 1644 920,clip,width=\textwidth]{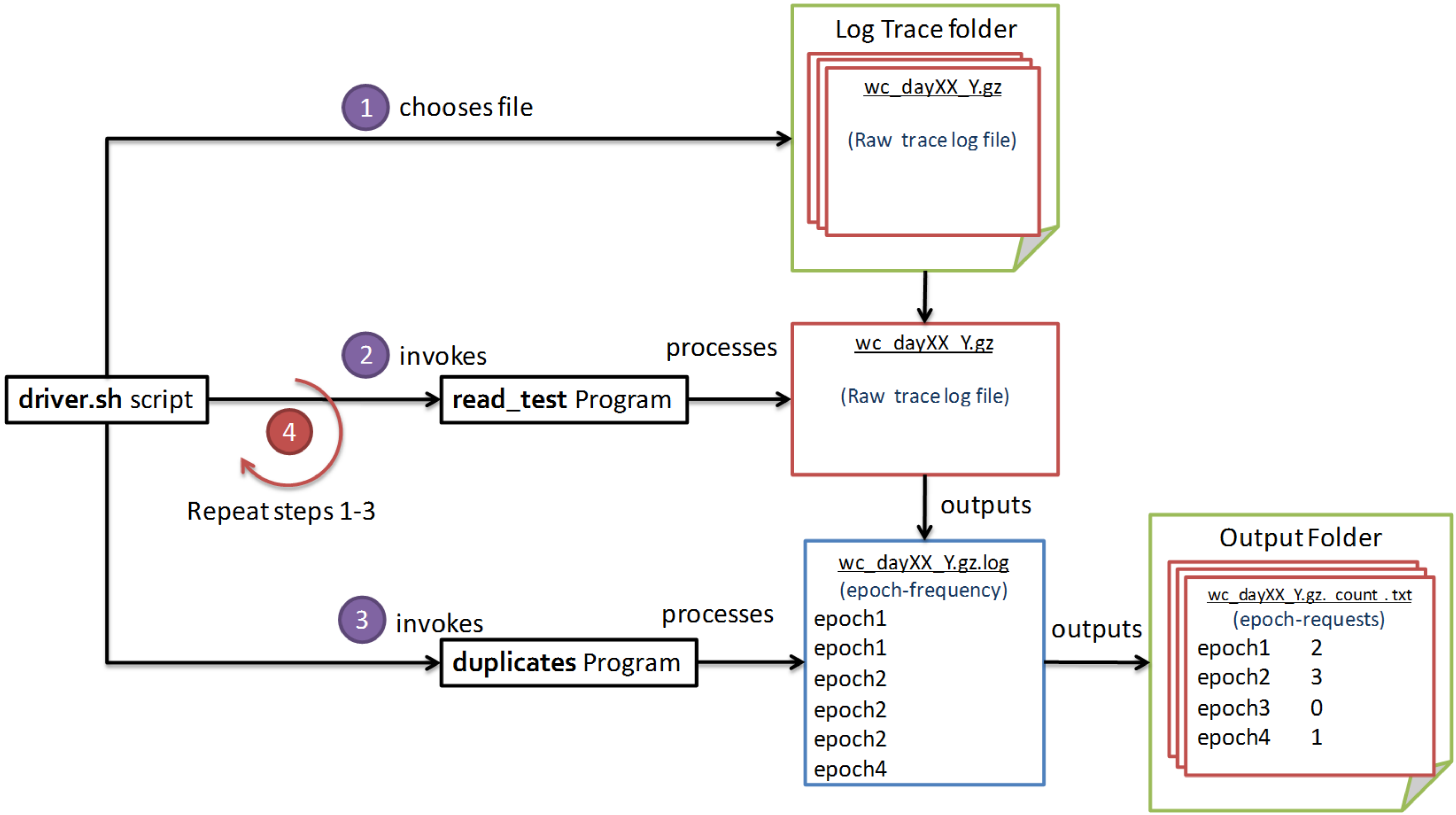}
\caption{Process Seconds Data}
\label{fig:ProcessSecondsData}
\end{figure}

Obtaining the \textit{epoch-requests} data set required further processing steps, therefore a \texttt{driver.sh} shell script and two programs: \texttt{read\_test} and \texttt{duplicates} were custom written to automate the process. The \texttt{read\_test} program was developed by modifying the \texttt{read} program --- needing single line source addition --- to process each single log trace file, output the epoch of each request and generate the intermediary data set files: \textit{epoch-frequency}. Each of these data files contained epochs occurring zero or multiple times, e.g. if an epoch occurred 10 times in this file then 10 requests were received during that particular epoch (refer \figurename~\ref{fig:ProcessSecondsData}). The \texttt{duplicates} program was written to process the \textit{epoch-frequency} data and generate the \textit{epoch-requests} data set. The program simply noted the epoch integer and the number of times they occurred, outputting each epoch and the associated frequency, thereby producing a condensed data set. To begin the whole process, \texttt{driver.sh} was run which invoked the \texttt{read\_test} and \texttt{duplicates} programs to automatically retrieve \textit{epoch-requests} for all the files in the Log Trace folder. \figurename~\ref{fig:ProcessSecondsData} and the following describes the steps that \texttt{driver.sh} script performs when run:

\begin{description}
	\item{Step 1:}	\textbf{Chooses} one file from Log Trace folder. Chosen File: \texttt{wc\_dayXX\_Y.gz}
	\item{Step 2:}	\textbf{Invokes} \texttt{read\_test} program on the chosen file to extract \textit{epoch-frequency} data. Output: \texttt{wc\_dayXX\_Y.gz.log}.
	\item{Step 3:}	\textbf{Invokes} \texttt{duplicates} program to generate \textit{epoch-requests} data.\\
	 Output: \texttt{wc\_dayXX\_Y.gz.count.txt}
	\item{Step 4:}	\textbf{Repeat} steps 1-3 above until no files remains to be processed.
\end{description}

\subsection{Data Format}
\label{subsec:dataFormat}
This sub-section describes briefly how data in the \textit{day-requests} and \textit{epoch-requests} data sets are organized. Based on the data collection process as discussed in section \ref{subsec:dataCollection} the \textit{day-requests} data set contains the day and corresponding requests received that day, labeled as `DAY' and `REQUESTS' columns respectively. Also added manually to the data set are the number of matches that were played for each day labelled as `MATCHES'. Another column  `ISMATCH' was added to indicate if a match was played on that day, if so then 1 was used as the value and 0 otherwise. For example, if two matches were played on a particular day, then `ISMATCH' would have the value 1 and the `MATCHES' would be set to 2 for that day. The data for the number of matches for each day was obtained from \cite{2012_fifaMatches}, however it was later found that FIFA Archive also includes the information \cite{1998_FIFAArchive}. Following shows a short sample of how \textit{day-requests} data set is organized:
\begin{verbatim}
DAY	REQUESTS	MATCHES	ISMATCH
...
45	20068724	0	0
46	50395084	2	1
47	52406319	2	1
48	48956621	3	1
49	23528986	3	1
50	21093494	3	1
51	58013849	3	1
52	40732114	2	1
...
\end{verbatim}

Based on the data collection process as discussed in section \ref{subsec:dataCollection} the \textit{epoch-requests} data set contains the epoch and corresponding requests --- i.e., requests/second --- received during that epoch labeled as `EPOCH' and `REQUESTS' columns respectively. Following shows a sample of how \textit{epoch-requests} data set is organized:
\begin{verbatim}
EPOCH	REQUESTS
898207201	145
898207202	242
898207203	276
898207204	283
898207205	285
...
\end{verbatim}

\section{ANN and Time-series forecasting}
\label{sec:ANN_MATLAB}
This section begins with a brief introduction to ANN, followed by discussion of ANN as a method for time-series forecasting \cite{2005_TimeSeriesForecasting}. The tools provided by MATLAB that aid in ANN time-series forecasting are also mentioned.

\subsection{ANN}
\begin{figure}[h]
\centering
\includegraphics[keepaspectratio,viewport=0 0 946 578,clip,width=0.55\textwidth]{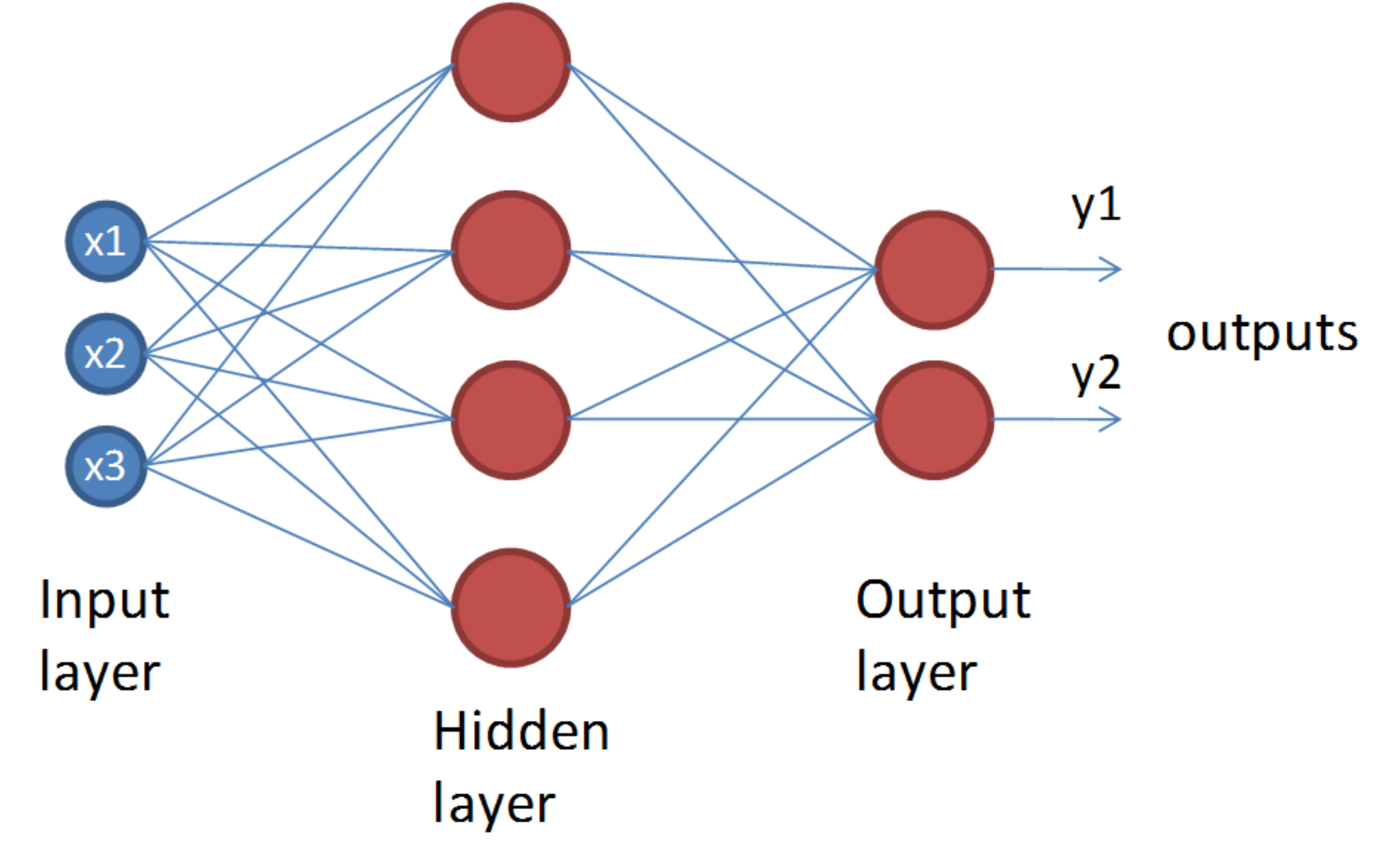}
\caption{ANN model showing the input, hidden and output layers. This is an example of feedforward neural networks where the output generated from a layer is fed as input to the next layer \cite{2001_NNShortTermLoadForecast}. }
\label{fig:ANN}
\end{figure}
The concept of artificial neural networks (ANN) is derived from the human nervous system, where a network of neurons, i.e. neural network, processes signals \cite{2010_IdentPredictInternetTraffic,2001_NNShortTermLoadForecast,1998_ForecastingANN}. Each neuron in ANN receives inputs, and processes them based on mathematical functions and relations, generating an output that is either fed as input to another neuron or served as the output of the whole ANN (refer \figurename~\ref{fig:ANN}). The layer comprising the input signals is known as input layer, the last layer of neurons that generate the final outputs is the output layer and the layers between input and output layers are known as hidden layers. In the case of multi-layer perceptron (MLP), a.k.a. feedforward neural networks, the output generated from a layer is fed as input to the next layer  \cite{2001_NNShortTermLoadForecast,2010_IdentPredictInternetTraffic}. The connections between neurons have weights, which affect the output of the network. Other factors that affect the output are the bias value of the neuron and the transfer function \textit{f}, which are both explained through the equation below. The output of neuron \textit{i} with \textit{R} inputs is described by the equation \cite[p.~1-7]{2012_MatlabNNUG}:
\[
output = f(\sum_{j=1}^R (weight_{i,j} * input_j)\ +\ bias)
\]

There could be various transfer functions such as sigmoid (tansig) and linear (purelin) \cite{2012_MatlabNNUG}. For our purposes we use sigmoid function for hidden layers and the linear function is used for the output layer. Sigmoid transfer function is described as follows \cite{2010_IdentPredictInternetTraffic}:
\[
f(x) = \frac{1}{1 + e^{-x}}
\]

Before using the neural network, training data comprising of rows of inputs and desired outputs are fed to the neural network for training purposes, which adjusts the weights of the connections based on a selected training algorithm. Back propagation \cite{2010_IdentPredictInternetTraffic} is a well-known training algorithm. For the purposes in this article, Levenberg-Marquardt backpropagation \cite{2012_MatlabNNUG} training algorithm has been used. Once the network it trained, test inputs are fed to the network and the network is simulated to produce the outputs.

\subsection{Time-series forecasting and MATLAB}
ANN models are useful for Time-series forecasting \cite{2005_TimeSeriesForecasting,2012_MatlabNNUG}, where the future outcome of a variable is predicted through the use of current and previous time values of the variables. ANN have been applied in forecast relating to finance and markets, electric power load, sunspots, temperature of environment, airline passengers, etc. \cite{1998_ForecastingANN}. Interested readers may refer to \cite[pp.~1-6--1-7]{2012_MatlabNNGS} and \cite[pp.~39--40]{1998_ForecastingANN} for further applications of ANN. 

In general, for time-series forecasting the output $y(t)$ is predicted based on previous $d$ delayed inputs. An example is nonlinear auto autoregressive (NAR) prediction provided by MATLAB which is based on the following equation \cite{2012_MatlabNNGS}:

\[
y(t) = f(y(t-1), \ldots, y(t-d))
\]

The mean squared error (MSE), which is a method for determining the error of prediction, is calculated as follows based on $N$ predictions, the target input values $t_i$ and the predicted values $a_i$ \cite[p.~2-16]{2012_MatlabNNUG}:
\[
MSE = \frac{1}{N}\sum\limits_{i=1}^{N}{ (t_{i} - a_{i})^2 }
\]

\begin{figure}[h]
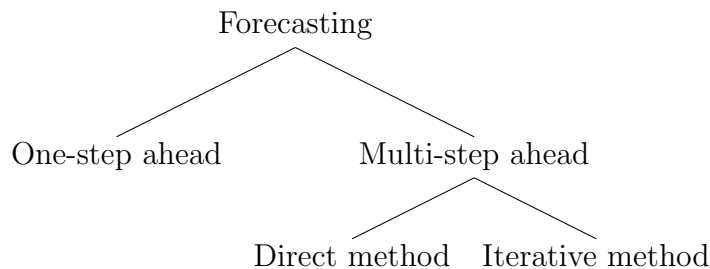

\centering
\Tree [.{Forecasting} {One-step ahead} [.{Multi-step ahead} {Direct method} {Iterative method} ] ]
\caption{Forecasting can be subdivided into one-step ahead or multi-step ahead forecasting \cite{1998_ForecastingANN}. Multi-step ahead could be further classified into direct and iterative forecasting \cite{1998_ForecastingANN}.}
\label{fig:Forecasting}
\end{figure}

\figurename~\ref{fig:Forecasting} shows the different types of forecasting. Forecasting can be subdivided into one-step ahead or multi-step ahead forecasting \cite{1998_ForecastingANN}. One-step ahead forecasting generates a single output predicting the value of y for the next time-step only, whereas multi-step ahead forecasting predicts value of $y$ for $m$ future time-steps, i.e. $y(t), y(t+1), \ldots, y(t + m - 1)$. Multi-step forecasting could further be classified into direct and iterative forecasting approaches \cite{1998_ForecastingANN}. In direct approach, there are multiple output nodes, whereas in iterative method, a single output is looped back as input to iteratively predict future values \cite{1998_ForecastingANN}. In this article, the main focus is on one-step ahead forecasting, however, a simple multi-step prediction has also been performed.

\begin{figure}[h]
\centering
\resizebox{0.85\textwidth}{!}{%
\Tree [.{MATLAB Time Series Tool\\(ntstool)} {Non-linear Autoregressive \\(NAR) network} {Non-linear Autoregressive with exogenous input \\(NARX) network} {Non-linear Input Output\\ (NIO) network} ]
}
\caption{nstool}
\label{fig:ntstool}
\end{figure}

In this article, MATLAB is used as a tool for time-series forecasting. To help with the forecasting and with ANNs in general, MATLAB provides the Neural Network Toolbox \cite{2012_MatlabNNGS,2012_MatlabNNUG}. ANN can easily be created, trained and simulated through the toolbox. Alongside, graphing options to show error, performance and response is available. For using the toolbox one can feed commands to the command-line (also by running scripts) or use GUI-based aids \cite{2012_MatlabNNGS}. The focused time-delay neural network (FTDNN) \cite{2012_MatlabNNUG} is a simple time-series prediction network that can be created through the command-line by calling the command \texttt{timedelaynet}. This is same as the one-step ahead prediction described above. For GUI options, the Time Series Tool (\texttt{ntstool}) \cite{2012_MatlabNNGS} graphical interface (\figurename~\ref{fig:ntstool}) provides users the option to choose from different predefined ANN: Non-linear Autoregressive (NAR), Non-linear Autoregressive with exogenous input (NARX) and Non-linear Input Output  (NIO) networks. NAR and NARX network are networks where output is fed-back into the network for prediction purposes, however, during training the feedback loop could be left open as original inputs are available and closed later for simulation \cite{2012_MatlabNNUG}. The main distinction between NAR and NARX networks is the latter not only uses the variable to be predicted as inputs but also another set of delayed inputs: $x(t), x(t-1), \ldots, x(t-d) $ for the forecast of $y(t)$ \cite{2012_MatlabNNUG}. Finally, the NIO network only uses $x(t), x(t-1) , \ldots, x(t-d)$ inputs for prediction of $y(t)$. The GUI allows saving the actions performed --- including network creation, training and testing --- as an auto-generated script which can either be modified or directly run through the command line.

\begin{figure}[h]
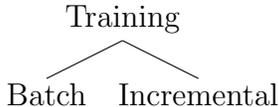

\centering
\Tree [.{Training} {Batch} {Incremental} ]
\caption{Batch and Incremental training}
\label{fig:Training}
\end{figure}

Networks can be trained in two ways (\figurename~\ref{fig:Training}). If batch training used is then the weights and biases of the network are updated when all of the input-output rows have been provided to the network, whereas, in the case of incremental training, each input-output row updates the network \cite{2012_MatlabNNUG}. In this article, FTDNN and batch training are mostly used for time-series forecasting; however, incremental training and NARX network have also been tried in different cases, details of which are available in the next section.

\section{Method}
\label{sec:Method}
This section describes the method followed to model the 1998 World Cup website requests using ANN. The aim is to evaluate the performance of different networks and determine how well the requests rate were predicted through the models. To begin, the first step was data collection, which has been described is detail in Section \ref{subsec:dataCollection}. Once data was available as two data sets: \textit{day-requests} and \textit{epoch-requests}, the next step required MATLAB to create ANN, train them and finally simulate the network to perform the prediction. For this purpose, multiple MATLAB scripts were used to automate some portions of the process. Through the use of \texttt{ntstool}, few scripts were initially auto-generated, which were then modified for the purposes described here, while the other scripts were custom written. The following four scripts were used:
\begin{enumerate}
  \item	\texttt{createNetwork.m}: create the ANN based on chosen network structure.
  \item	\texttt{trainSimNetwork.m}: train and simulate the ANN.
  \item \texttt{initNetwork.m}: initalize network weights. This is useful to begin with weights that might be trained to reach a better performance.
  \item	\texttt{revertNetwork.m}: revert to the previous network weights just before \texttt{initNetwork.m} was called. This is useful if after calling \texttt{initNetwork.m} and then training, the network showed poor performance and in which case the previous network weights were better.
\end{enumerate}

\begin{figure}[h!t!b]
\centering
\includegraphics[keepaspectratio,viewport=0 0 1260 381,clip,width=\textwidth]{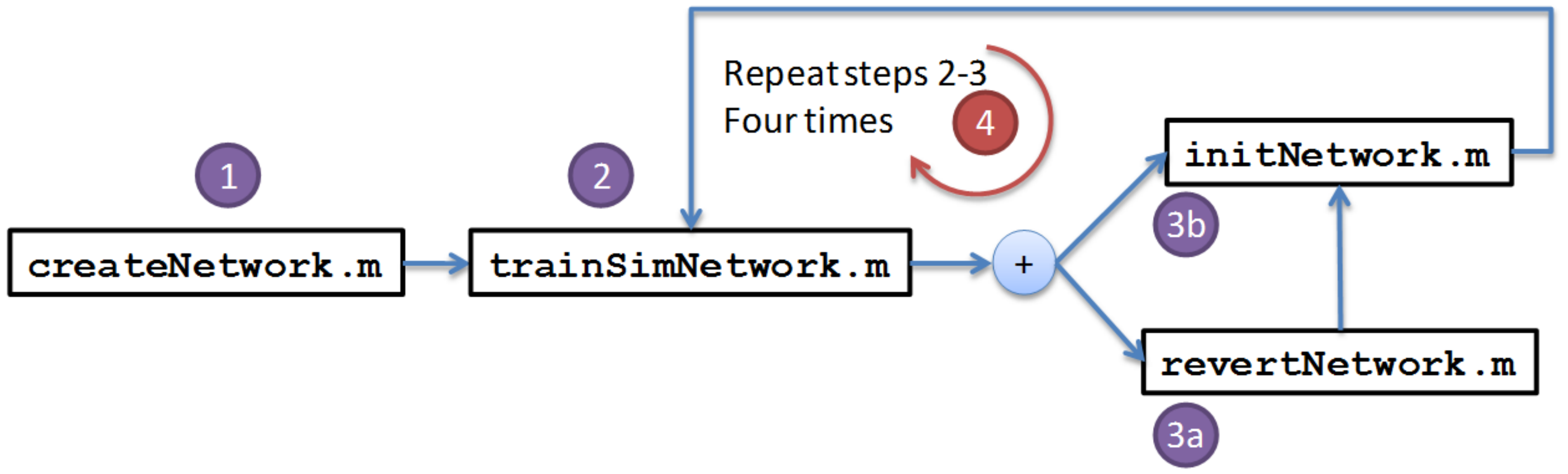}
\caption{Custom MATLAB Scripts: Process Flow Diagram}
\label{fig:MATLABScriptsProcessFlowDiagram}
\end{figure}

Once the scripts were developed the following process was manually followed (\figurename~\ref{fig:MATLABScriptsProcessFlowDiagram}):
\begin{description}
	\item{Step 1:}	\textbf{Execute} \texttt{createNetwork.m} script.
	\item{Step 2:}	\textbf{Execute} \texttt{trainSimNetwork.m} script. Go to step3b (\texttt{initNetwork.m}) if performance is better than previous network, else goto step 3a (\texttt{revertNetwork.m}).
	\item{Step 3a:}	\textbf{Execute} \texttt{revertNetwork.m} script.
	\item{Step 3b:}	\textbf{Execute} \texttt{initNetwork.m} script.
	\item{Step 4:}	\textbf{Repeat} steps 2-3 above for another four times.
\end{description}

To evaluate different networks, 13 cases were tried and their results analyzed. The process as described above using the MATLAB scripts were followed for each case with minor script modifications as per the requirements. \tablename~\ref{tab:ListOfCases} lists the 13 cases. D1--D11 use the \textit{day-requests} data set and S12--S13 use the \textit{epoch-requests}. D1 describes the base network and subsequent cases include modifications to this network's structure, data distribution (training/validation/test), training mode and/or input delays. D1 uses a FTDNN to perform one-step ahead prediction and has two input delays. The data is trained using batch mode and the contiguous distribution for training/validation/testing is 70\%/15\%/15\%. All cases except D9-D11 use open-loop models, whereas D9-D11 use closed-loop networks for simulation, although their training is performed using open-loop network. D9 uses NAR network and D10 and D11 use NARX network. The $x(t)$ input for D10 and D11 are `MATCHES' and `ISMATCH' columns of the data set, respectively. The \textit{epoch-requests} data set is modeled by S12, which is sub-divided into cases S12a and S12b. S12a models the first 1000 rows of \texttt{wc\_day6\_1.gz.count.txt} file and 12b using the same network simulates the whole \texttt{wc\_day6\_1.gz.count.txt} file. The last one, case 13, models the beginning 1000 points of \texttt{wc\_day66\_10.gz.count.txt} data set.

\begin{longtable}{|l|p{0.575\textwidth}|}\caption{A List of cases that have been tested using MATLAB Neural Network Toolbox.}{\label{tab:ListOfCases}} \\
\hline
\textbf{Case} & \textbf{Description} \\
\hline
\endfirsthead

\hline
\textbf{Case} & \textbf{Description} \\
\hline
\endhead

\multicolumn{2}{r}{\textit{Continued on next page}} \\
\endfoot
\hline
\endlastfoot

D1 & 
\begin{inparaenum}[i)]
\item FTDNN
\item Batch training mode
\item \textit{day-requests} data (92 points)
\item training/validation/testing distribution: 70\%/15\%/15\% (contiguous data)
\item Delays = 1:2, i.e. inputs = y(t-1) and y(t-2)
\item one-step ahead prediction
\item hiddenLayerSize = 10
\item one hidden layer
\end{inparaenum}
\\
\hline
D2 & Same as D1 except: hiddenLayerSize = 30 \\ 
\hline
D3 & Same as D1 except: two hidden layers \\
\hline
D4 & Same as D1 except: training/validation/testing distribution: 80\%/10\%/10\%\\
\hline
D5 & Same as D1 except: training/validation/testing distribution: 60\%/20\%/20\%\\
\hline
D6 & Same as D1 except: hiddenLayerSize = 1\\
\hline
D7 & Same as D1 except: Incremental (adapt) training mode\\
\hline
D8 & Same as D1 except: Delays = 1:7 \\
\hline
D9\footnote{Closed-loop}
	& Same as D1 except: NAR network. Delays = 2:3, i.e. inputs = y(t-2) and y(t-3). 2-step ahead prediction.  (closed-loop) \\
\hline
D10\footnotemark[\value{footnote}]
	& Same as D1 except: NARX network. Exogenous input is `MATCHES'. Uses open-loop for training and closed-loop for simulation.\\
\hline
D11\footnotemark[\value{footnote}]
	& Same as D1 except: NARX network. Exogenous input is `ISMATCH'. Uses open-loop for training and closed-loop for simulation.\\
\hline
S12a & Same as D1 except: Data is Seconds data beginning 1000 points of \texttt{wc\_day6\_1.gz.count.txt} epoch-request data file .\\
\hline
S12b\footnote{Simulation only. Uses network trained from S12a.} 
	& Same as S12a except Simulation only. Network from S12a is used for simulation . The data contains all points of \texttt{wc\_day66\_10.gz.count.txt} \textit{epoch-requests} file \\
\hline
S13 & Same as S12a except: Data is Seconds data beginning 1000 points of \texttt{wc\_day66\_10.gz.count.txt} epoch-request file .
\\
\end{longtable}

After trying each case, the results which include mean-squared error (MSE) and the correlation coefficient \textit{R} of the of the modeled data. These results are analyzed in the following section.

\section{Results}
\label{sec:Results}
\begin{figure}[h]
\centering
\includegraphics[keepaspectratio,viewport=0 0 1712 900,clip,width=\textwidth]{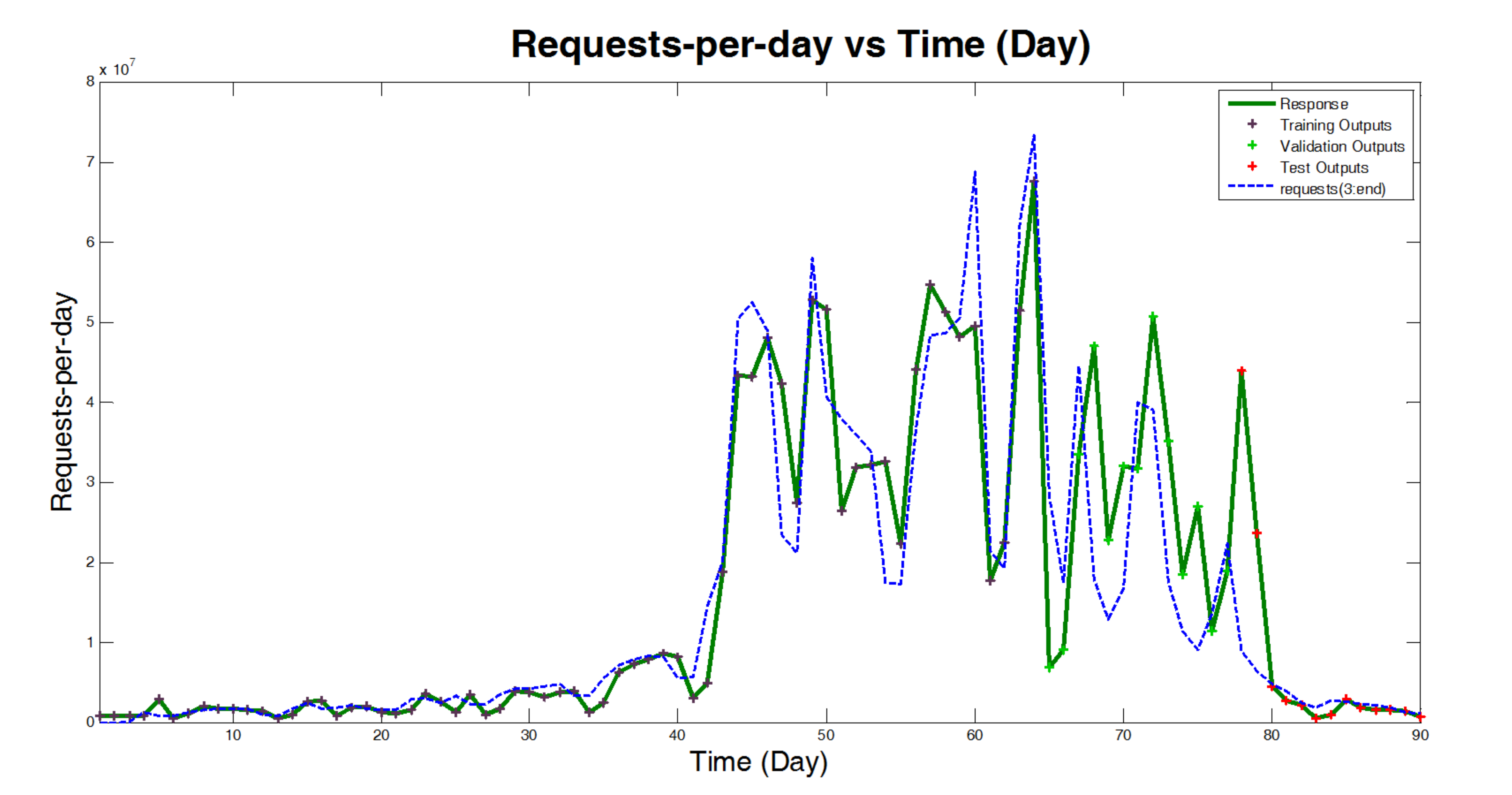}
\caption{Case D1 - Requests-per-day vs. Day. Day 1 value on the graph corresponds to the first prediction and since there are 1:2 delays then Day 1 here corresponds to actual data set's Day 3 (Day 1 and Day 2 values of actual data set as delayed inputs).}
\label{fig:matlab-dayMatch-1-article}
\end{figure}

\begin{figure}[h]
\centering
\includegraphics[keepaspectratio,viewport=0 0 1712 900,clip,width=\textwidth]{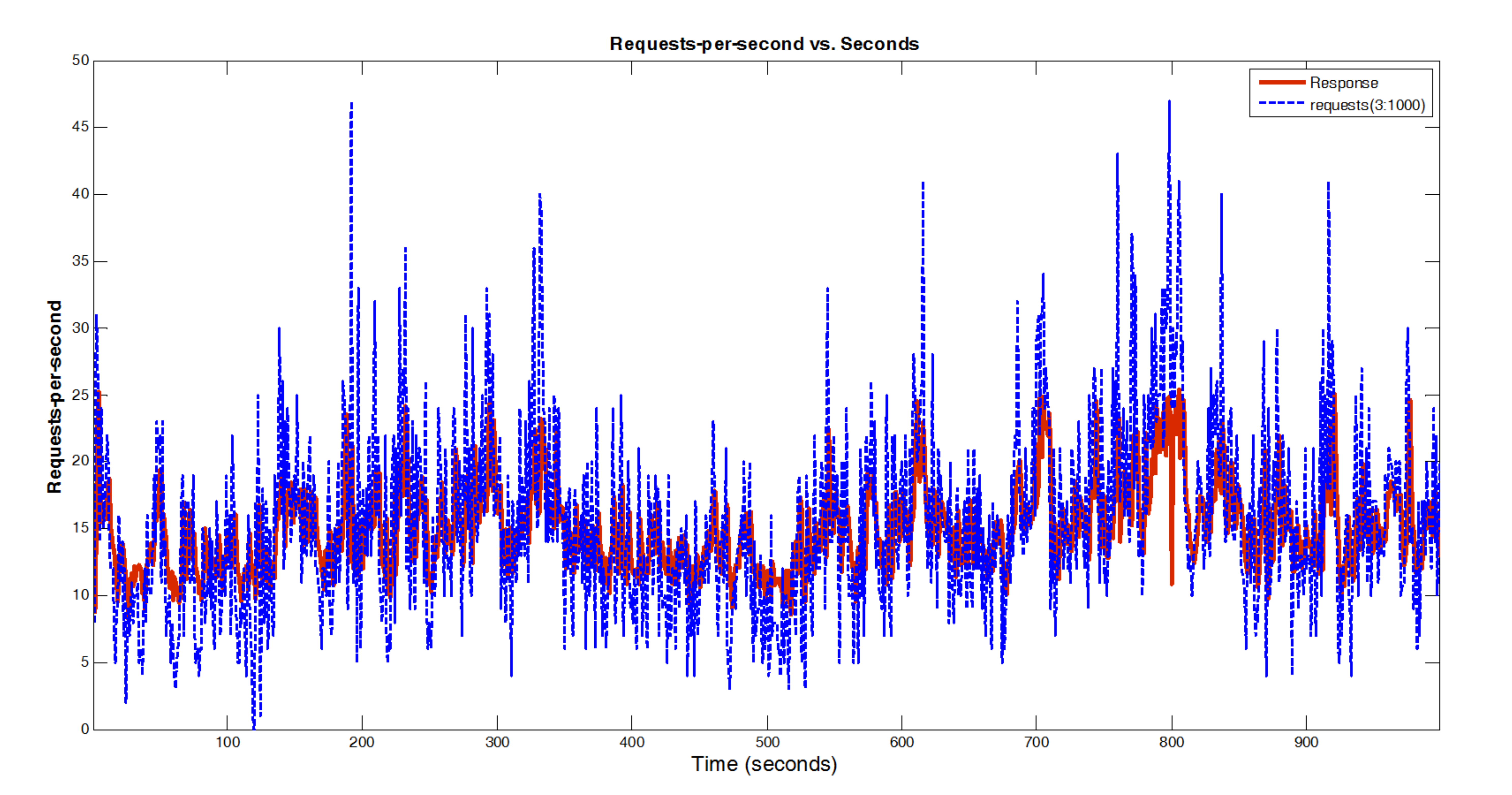}
\caption{Case S12a - Requests-per-second vs. Seconds of first 1000 points of day 6. Second 1 value on this graph corresponds to the first prediction and since there are 1:2 delays then Second 1 here corresponds to actual data set's Second 3, where Second 1 and Second 2 values of actual data set are the delayed inputs.}
\label{fig:matlab-dayMatch-12-1000-article}
\end{figure}

\begin{figure}[h]
\centering
\includegraphics[keepaspectratio,width=\textwidth]{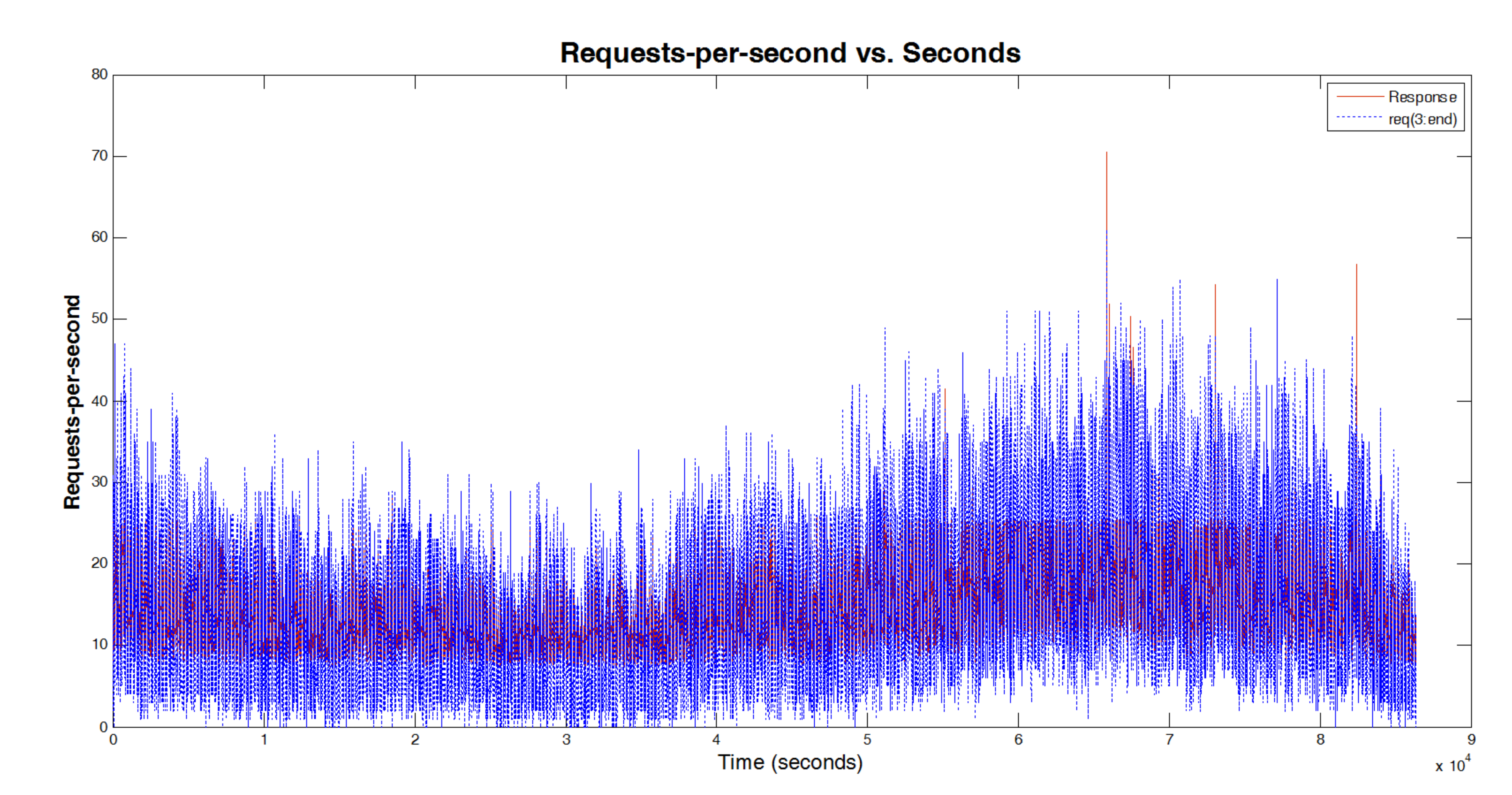}
\caption{Case S12b - Requests-per-second vs. Seconds of complete day 6. Second 1 value on this graph corresponds to the first prediction and since there are 1:2 delays then Second 1 here corresponds to actual data set's Second 3, where Second 1 and Second 2 values of actual data set are the delayed inputs.}
\label{fig:matlab-dayMatch-12b-complete-article}
\end{figure}

\begin{figure}[h]
\centering
\includegraphics[keepaspectratio,width=\textwidth]{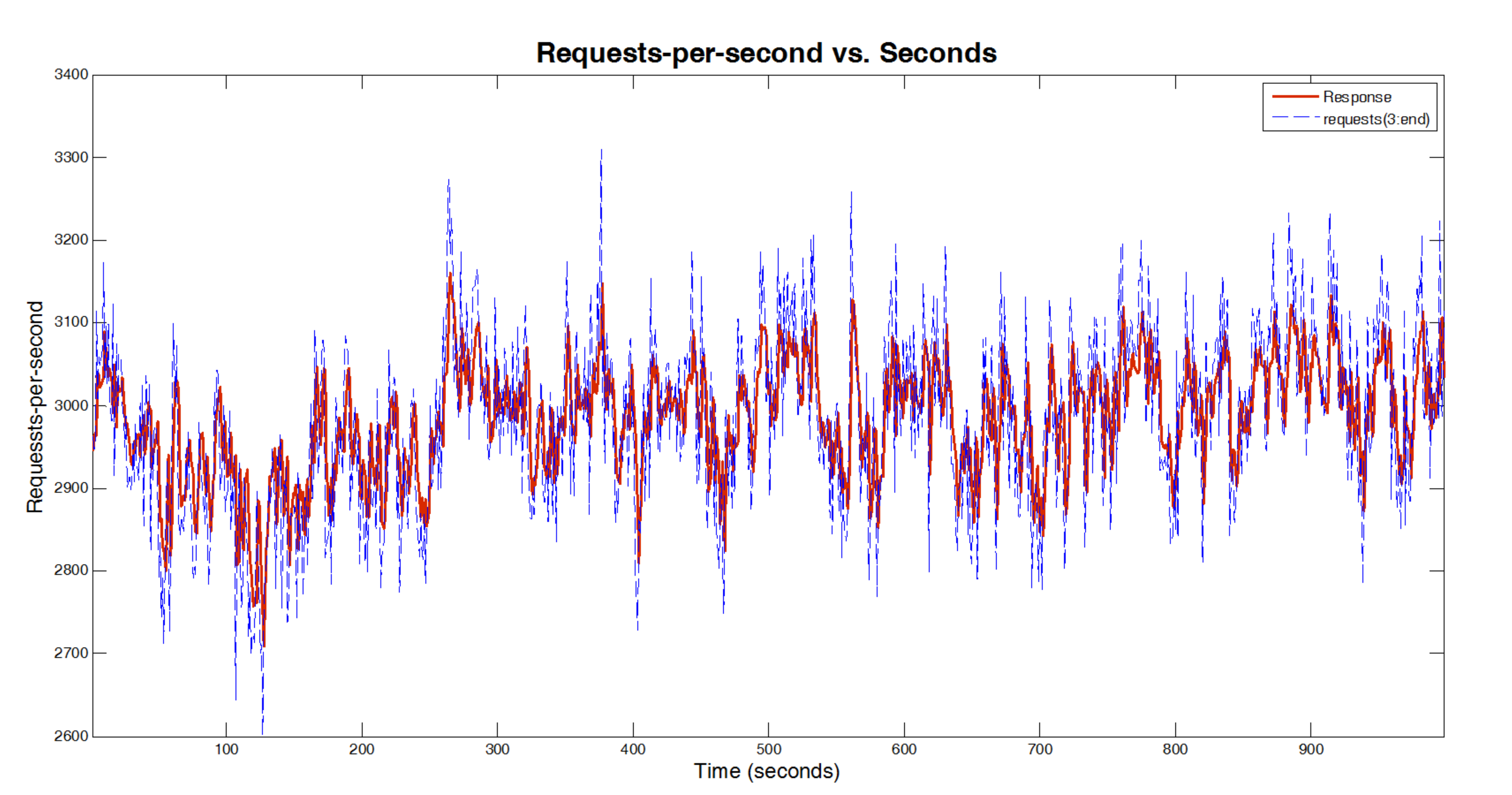}
\caption{Case S13 - Requests-per-second vs. Seconds of first 1000 points of day 66-part10. Second 1 value on this graph corresponds to the first prediction and since there are 1:2 delays then Second 1 here corresponds to actual data set's Second 3, where Second 1 and Second 2 values of actual data set are the delayed inputs.}
\label{fig:matlab-dayMatch-13-1000-article}
\end{figure}

In this section, the results from studying the 13 cases --- described earlier in section \ref{sec:Method} --- are presented. The main results are the complete performance in MSE of the network (for all training, validation and testing) and the correlation coefficient \textit{R} (\tablename~\ref{tab:ResultsCases}). Four figures (\figurename~\ref{fig:matlab-dayMatch-1-article}, \figurename~\ref{fig:matlab-dayMatch-12-1000-article}, \figurename~\ref{fig:matlab-dayMatch-12b-complete-article}, \figurename~\ref{fig:matlab-dayMatch-13-1000-article}), which graphically show the requests vs. day and requests vs. seconds graphs for cases D1, S12a, S12b and S13 respectively, are also discussed.

In \figurename~\ref{fig:matlab-dayMatch-1-article} the x-axis represents the predicted time steps in days, therefore, day 1 here corresponds to first predicted day based on 2 delayed inputs, as case D1 uses two delayed inputs. To clarify further, for D1, the ``actual'' data set's day 1 and day 2 are the delayed inputs and the first actual predicted day is day 3, i.e. day 1 in the graph corresponds to day 3 of the data set. The dashed blue line indicated the actual requests/day values (i.e. targets) and the solid green line is the response by simulating the neural networks. The first contiguous 70\% of the data, shown in purple cross points, are the training data, the next 15\% in blue cross points are the validation data and the remaining in red are the testing data. From the graph it is seen that that the ANN has been able to model the requests/day reasonably well. The MSE is 0.0127 and \textit{R} is 0.90134, the latter showing that there is strong relation between the targets and the response. 

In \figurename~\ref{fig:matlab-dayMatch-12-1000-article} the x-axis displays the predicted time steps in seconds, therefore, second 1 here corresponds to first predicted second based on 2 delayed inputs, following the same explanation as for graph in \figurename~\ref{fig:matlab-dayMatch-1-article}. In the graph one-step ahead prediction has been used. The target requests/seconds are shown in dashed blue line and the solid red line represents the outputs. From the graph the ANN shows to model the requests/second to a reasonably good degree of accuracy along with a network performance of 0.0163. Correlation coefficient \textit{R} = 0.5043, depicting a positive but not a strong correlation between the targets and the response.

Similarly, \figurename~\ref{fig:matlab-dayMatch-12b-complete-article} shows the simulation from using the same network as obtained from case S12a. The MSE for case 12b is 0.0151, which is comparatively better than case 12b.

\figurename~\ref{fig:matlab-dayMatch-13-1000-article} shows the graph for the first 1000 seconds of day 66-part10. The MSE for case 13 is 0.0125, and the correlation coefficient \textit{R} = 0.65686, overall depicting a reasonable prediction of the requests/second for day 66-part10.

\begin{longtable}{|l|p{4.5cm}|l|l|l|}
\caption{Results of the 13 cases}{\label{tab:ResultsCases}} \\
\hline
\textbf{Case} & \textbf{Description} & \textbf{Complete Perf. (MSE)} & \textbf{R} \\
\hline
\endfirsthead

\hline
\textbf{Case} & \textbf{Description} & \textbf{Complete Perf. (MSE) } & \textbf{R} \\
\hline
\endhead

\multicolumn{2}{r}{\textit{Continued on next page}} \\
\endfoot
\hline
\endlastfoot

D1						& Base network using \textit{day-requests}						& 0.0127		&0.90134\\
\hline
D2						& hiddenLayerSize = 30									& 0.0295		&0.84082\\
\hline
D3						& two hidden layers										& 0.0087		&0.92925\\
\hline
D4						& 80\%/10\%/10\% (data)									& 0.0078		&0.93894\\
\hline
D5						& 60\%/20\%/20\% (data)									& 0.0252		&0.84021\\
\hline
D6						& hiddenLayerSize = 1									& 0.0188		&0.8457\\
\hline
D7						& Incremental training									& 0.0768		&0.57631\\
\hline
D8						& Delays = 1:7											& 0.0431		&0.80433\\
\hline
D9\footnote{Closed-loop}
						& NAR network. 2-step ahead prediction (closed-loop)	& 0.0682		&0.50645\\
\hline
D10\footnotemark[\value{footnote}]
						& NARX network. Exogenous input: `MATCHES' (closed-loop)	& 0.0779		&0.63225\\
\hline
D11\footnotemark[\value{footnote}]
						& NARX network. Exogenous input: `ISMATCH'	(closed-loop)	& 0.0498		&0.54486\\
\hline
S12a					& epoch-seconds data set (1000-points of day 6)			& 0.0163		&0.5043\\
\hline
S12b\footnote{Simulation only. Uses network trained from S12a.}
						& Network from S12a. Complete day 6 data. 				& 0.0151		&0.61852\\
\hline
S13						& epoch-seconds data set (1000-points of day 66\_10)	& 0.0125		&0.65686\\
\end{longtable}

\tablename~\ref{tab:ResultsCases} summarizes the results of the networks evaluation. Here, the result discussions are presented. Amongst the \textit{day-requests} data set cases, the best performers are D3, D4 and D1 with performance range between 0.0078--0.0127 and correlation coefficient range between 0.90134--0.93894. In particular, case D4 network has the highest performance. D4 uses 80\% of the data for training, a possible reason for having better results than others. Using two hidden layers (case D3) network results in better performance than using one hidden-layer (base case D1). 

Cases D2, D5, D6 and D8 show average performance with MSE ranging between 0.0188--0.0431 and correlation coefficient between 0.80433--0.8457. Having 30 neurons or 1 neuron in the hidden layer --- as in cases D2 and D6, respectively ---  instead of using 10 neurons (base case D1), does not help increase the performance and on the other hand causes a decrease. From the results it is seen that using more data for training provides better performance. This is evident from case D4, which uses 80\% data for training, show better results in comparison to D1, using 70\% data for training, which in turn has a better performance in comparison to D5 using 60\% of data for training. With two delays (base case D1) the prediction shows better performance in comparison to seven delay inputs (case D8).

Cases D7, D9, D10 and D11 are the worst performers with MSE ranging between 0.0498--0.0779 and correlation coefficient between 0.50645--0.63225. The results show that closed-loop NARX network (cases D10 and D11) which have exogenous input of `ISMATCH' or `MATCHES' show poor performance and low correlation with targets and response. Furthermore, iterative two-step ahead prediction (case D9) also doesn't not indicate good performance in comparison to one-step ahead performance (case D1). Alongside, incremental training (case D7) does not fare well in comparison to batch training.

Amongst the \textit{epoch-requests} data set cases, S12a and S12b model 1000-data points and complete data-points of day 6 requests, respectively. S12a and S12b network have a performance of 0.0163 and 0.0151, respectively. The network trained for S12a shows comparatively better performance when used to predict the data of S12b than predicting data of 12a. Day 66 data (1000-points) is modeled by network in case S13 and has a MSE of 0.0125 and correlation of 0.65686.

Summarizing the results above, it is seen that if more percentage of data is used for training then the network is able to perform a better prediction. Two hidden-layers shows better results than one hidden layers based on the data used. Also using too few or too many neurons decreases performance and in particular, based on the data modeled, use of 10 neurons comparatively shows better performance. Batch training also performs better than incremental training. Furthermore, using exogenous inputs in the \textit{day-requests} data set did not help improve the performance. Finally, one-step ahead prediction shows better results than iterative multi-step ahead prediction. Based on the results --- when data distribution for training, validation and simulation is not varied and when batch training is employed --- a network with two-hidden layers and 10 hidden layer size shows the best performance.

\section{Conclusions}
\label{sec:Conclusions}
In this article, the workload intensity of FIFA World Cup website has been modeled by using ANN. Artificial neural networks have been employed for time-series prediction of two data sets: \textit{day-requests} and --- day 6 and day 66-part10 of --- \textit{epoch-requests}. In total, 13 cases have been studied and compared. One base network is used and subsequent cases include modifications to this network's structure, data distribution (training/validation/test), training mode and/or input delays. The method followed to collect and process data, and perform the experiments have been detailed in this article. The networks were created, trained and simulated using MATLAB Neural Network Toolbox. The results of all the cases have been presented, discussed, compared and summarized. Based on the results --- when data distribution for training, validation and simulation is not varied and when batch training is employed --- a network with two-hidden layers and 10 hidden layer size shows the best performance. This network has shown to model the requests intensity with reasonable accuracy, as seen from the MSE and correlation coefficient.

As a future work, the relationship between the website workload intensity and the audience of the matches could be found. On the same note, the prediction of the expected audience, using website workload intensity as an input, could be made by employing artificial neural networks. The popularity and rankings of teams could also serve as an input for forecasting of website request rate and the expected audience in upcoming matches. Study into how network structure could be modified --- or other means such as adding inputs --- to help improve incremental and multi-step ahead prediction would be fruitful. ANN models could also be compared with other linear and non-linear prediction approaches and the results analyzed. If website and audience data is collected over the years then a more accurate prediction appears possible. Such assumptions could be tested and verified. 

Through this article the applicability of ANN in modeling and forecasting of website workload intensity has been demonstrated, which is not restricted to FIFA World Cup website only. ANN modeling could also be used for other websites of sporting events (e.g. Super Bowl or Stanley Cup) --- and any website in general --- thereby again establishing the role of ANN in forecasting and expanding the horizons of their use.
\clearpage

\bibliographystyle{elsarticle-num}
\bibliography{article}

\begin{thebibliography}{10}
\expandafter\ifx\csname url\endcsname\relax
  \def\url#1{\texttt{#1}}\fi
\expandafter\ifx\csname urlprefix\endcsname\relax\def\urlprefix{URL }\fi
\expandafter\ifx\csname href\endcsname\relax
  \def\href#1#2{#2} \def\path#1{#1}\fi

\bibitem{2011_YasirShoaibMAScThesis}
Y.~Shoaib, Performance measurement and analytic modeling of a web application,
  Master's thesis, Ryerson University, Toronto, ON (2011).

\bibitem{1984_QSP}
E.~D. Lazowska, J.~Zahorjan, G.~S. Graham, K.~C. Sevcik, Quantitative system
  performance: computer system analysis using queueing network models,
  Prentice-Hall, Inc., Upper Saddle River, NJ, USA, 1984.

\bibitem{WorldCup98Traces}
M.~Arlitt, T.~Jin,
  \href{http://ita.ee.lbl.gov/html/contrib/WorldCup.html}{{1998 World Cup Web
  Site Access Logs}}, [Accessed August 27, 2013] (August 1998).
\newline\urlprefix\url{http://ita.ee.lbl.gov/html/contrib/WorldCup.html}

\bibitem{LogTraces}
{The Internet Traffic Archive},
  \href{http://ita.ee.lbl.gov/html/traces.html}{Traces available in the
  internet traffic archive}, [Accessed August 27, 2013] (2008).
\newline\urlprefix\url{http://ita.ee.lbl.gov/html/traces.html}

\bibitem{NASAHTTP}
J.~Dumoulin, M.~Arlitt, C.~Williamson,
  \href{http://ita.ee.lbl.gov/html/contrib/NASA-HTTP.html}{{NASA-HTTP - Two
  Months of HTTP Logs from the KSC-NASA WWW Server}}, [Accessed August 27,
  2013] (1995).
\newline\urlprefix\url{http://ita.ee.lbl.gov/html/contrib/NASA-HTTP.html}

\bibitem{EPAHTTP}
L.~Bottomley, \href{http://ita.ee.lbl.gov/html/contrib/EPA-HTTP.html}{{EPA-HTTP
  - A Day of HTTP Logs from the EPA WWW Server}}, [Accessed August 27, 2013]
  (1995).
\newline\urlprefix\url{http://ita.ee.lbl.gov/html/contrib/EPA-HTTP.html}

\bibitem{1998_ForecastingANN}
G.~Zhang, B.~E. Patuwo, M.~Y. Hu,
  \href{http://www.sciencedirect.com/science/article/pii/S0169207097000447}{Forecasting
  with artificial neural networks:: The state of the art}, International
  Journal of Forecasting 14~(1) (1998) 35 -- 62.
\newblock \href {http://dx.doi.org/10.1016/S0169-2070(97)00044-7}
  {\path{doi:10.1016/S0169-2070(97)00044-7}}.
\newline\urlprefix\url{http://www.sciencedirect.com/science/article/pii/S0169207097000447}

\bibitem{2001_NNShortTermLoadForecast}
H.~Hippert, C.~Pedreira, R.~Souza, Neural networks for short-term load
  forecasting: a review and evaluation, Power Systems, IEEE Transactions on
  16~(1) (2001) 44 --55.
\newblock \href {http://dx.doi.org/10.1109/59.910780}
  {\path{doi:10.1109/59.910780}}.

\bibitem{2011_PredictionCloudNN}
J.~Prevost, K.~Nagothu, B.~Kelley, M.~Jamshidi, Prediction of cloud data center
  networks loads using stochastic and neural models, in: System of Systems
  Engineering (SoSE), 2011 6th International Conference on, 2011, pp. 276
  --281.
\newblock \href {http://dx.doi.org/10.1109/SYSOSE.2011.5966610}
  {\path{doi:10.1109/SYSOSE.2011.5966610}}.

\bibitem{2010_IdentPredictInternetTraffic}
S.~Chabaa, A.~Zeroual, J.~Antari, Identification and prediction of internet
  traffic using artificial neural networks, Journal of Intelligent Learning
  Systems and Applications 2~(3) (2010) 147--155.

\bibitem{2013_MatlabLM}
M.~H. Beale, M.~T. Hagan, H.~B. Demuth,
  \href{http://www.mathworks.com/help/toolbox/nnet/ref/trainlm.html}{{Levenberg-Marquardt
  backpropagation - MATLAB trainlm}}, {Matlab\textregistered}, release 2013a,
  [Accessed August 27, 2013] (2013).
\newline\urlprefix\url{http://www.mathworks.com/help/toolbox/nnet/ref/trainlm.html}

\bibitem{2011_LoadPredictionHotSpotDetection}
P.~Saripalli, G.~Kiran, R.~Shankar, H.~Narware, N.~Bindal, Load prediction and
  hot spot detection models for autonomic cloud computing, in: Utility and
  Cloud Computing (UCC), 2011 Fourth IEEE International Conference on, 2011,
  pp. 397--402.
\newblock \href {http://dx.doi.org/10.1109/UCC.2011.66}
  {\path{doi:10.1109/UCC.2011.66}}.

\bibitem{2013_NeuralWebServerWorkload}
T.~Van~Giang, D.~Vincent, B.~Seddik, Neural networks for web server workload
  forecasting, in: Industrial Technology (ICIT), 2013 IEEE International
  Conference on, 2013, pp. 1152--1156.
\newblock \href {http://dx.doi.org/10.1109/ICIT.2013.6505835}
  {\path{doi:10.1109/ICIT.2013.6505835}}.

\bibitem{2012_MatlabNNGS}
M.~H. Beale, M.~T. Hagan, H.~B. Demuth, {Neural Network Toolbox\texttrademark -
  Getting Started Guide}, {Matlab\textregistered}, {Version 7.0.3 (Release
  2012a)} (Mar. 2012).

\bibitem{2012_MatlabNNUG}
M.~H. Beale, M.~T. Hagan, H.~B. Demuth, {Neural Network Toolbox\texttrademark -
  User's Guide}, {Matlab\textregistered}, {Version 7.0.3 (Release 2012a)} (Mar.
  2012).

\bibitem{2011_timeManPage}
M.~Kerrisk,
  \href{http://www.kernel.org/doc/man-pages/online/pages/man2/time.2.html}{time(2)
  - linux manual page}, [Accessed August 27, 2013] (sep 2011).
\newline\urlprefix\url{http://www.kernel.org/doc/man-pages/online/pages/man2/time.2.html}

\bibitem{1998_FIFAArchive}
FIFA\textregistered,
  \href{http://www.fifa.com/worldcup/archive/edition=1013/index.html}{{FIFA.com
  - 1998 FIFA World Cup France \texttrademark}}, [Accessed August 27, 2013]
  (1994-2012).
\newline\urlprefix\url{http://www.fifa.com/worldcup/archive/edition=1013/index.html}

\bibitem{2000_1998WorldCupWorkloadCharacter}
M.~Arlitt, T.~Jin, A workload characterization study of the 1998 world cup web
  site, Network, IEEE 14~(3) (2000) 30 --37.
\newblock \href {http://dx.doi.org/10.1109/65.844498}
  {\path{doi:10.1109/65.844498}}.

\bibitem{1998_FranceDefeatBrazil}
B.~Macintyre, J.~Goodbody, G.~Gamini, France goes wild after 3-0 defeat of
  brazil, copyright - Copyright News International Newspapers Ltd. Jul 13,
  1998; Last updated - 2011-10-15 (Jul 13 1998).

\bibitem{2012_fifaMatches}
{Wikipedia, the free encyclopedia},
  \href{http://en.wikipedia.org/wiki/1998_FIFA_World_Cup}{1998 fifa world cup -
  wikipedia, the free encyclopedia}, [Accessed May 22, 2012] (may 2012).
\newline\urlprefix\url{http://en.wikipedia.org/wiki/1998_FIFA_World_Cup}

\bibitem{2005_TimeSeriesForecasting}
C.~Chatfield,
  \href{http://dx.doi.org/10.1111/j.1740-9713.2005.00117.x}{Time-series
  forecasting}, Significance 2~(3) (2005) 131--133.
\newblock \href {http://dx.doi.org/10.1111/j.1740-9713.2005.00117.x}
  {\path{doi:10.1111/j.1740-9713.2005.00117.x}}.
\newline\urlprefix\url{http://dx.doi.org/10.1111/j.1740-9713.2005.00117.x}

\end{thebibliography}

\end{document}